\documentclass{webofc}
\usepackage{graphicx} 
\graphicspath{{figures/}}
\usepackage{subcaption}
\usepackage[varg]{txfonts}
\usepackage[utf8]{inputenc}
\usepackage{newunicodechar}

\newunicodechar{∈}{$\in$}
\newunicodechar{π}{$\pi$}
\newunicodechar{θ}{$\theta$}
\usepackage[frozencache,cachedir=.]{minted}

\usepackage{natbib}
\usepackage{hyperref}
\usepackage{enumitem} 

\usepackage{lineno}

\title{Julia in HEP}

\author{\firstname{Graeme Andrew} \lastname{Stewart}\inst{1}\fnsep\thanks{\email{graeme.andrew.stewart@cern.ch}} \and
\firstname{Alexander} \lastname{Moreno Briceño}\inst{2} \and
\firstname{Philippe} \lastname{Gras}\inst{3} \and
\firstname{Benedikt} \lastname{Hegner}\inst{1} \and
\firstname{Uwe} \lastname{Hernandez Acosta}\inst{4,5} \and
\firstname{Tamas} \lastname{Gal}\inst{6} \and
\firstname{Jerry} \lastname{Ling}\inst{7} \and
\firstname{Pere} \lastname{Mato}\inst{1} \and
\firstname{Mikhail} \lastname{Mikhasenko}\inst{8} \and
\firstname{Oliver} \lastname{Schulz}\inst{9} \and
\firstname{Sam} \lastname{Skipsey}\inst{10}
}

\institute{EP-SFT, CERN, Geneva, Switzerland
\and
Universidad Antonio Nariño, Ibagué, Colombia
\and
IRFU-CEA, Université Paris-Saclay, Gif-sur-Yvette, France
\and
Center for Advanced Systems Understanding, Görlitz, Germany
\and
Helmholtz-Zentrum Dresden-Rossendorf, Dresden, Germany
\and
Erlangen Centre for Astroparticle Physics, Friedrich-Alexander-Universität, Erlangen, Germany
\and
Laboratory for Particle Physics and Cosmology, Harvard University, Cambridge MA, USA
\and
Ruhr Universität Bochum, Bochum, Germany
\and
Max-Planck-Institut für Physik, Munich, Germany
\and
School of Physics \& Astronomy, University of Glasgow, Glasgow, United Kingdom, G12 8QQ
}

\abstract{%
Julia is a mature general-purpose programming language, with a large ecosystem
of libraries and more than 12000 third-party packages, specifically
targeting scientific computing. Julia runs on x86, aarch64 and PowerPC architectures, and on all major GPU platforms.
As a language, Julia is as dynamic, interactive,
and accessible as Python with NumPy, but achieves run-time performance on par
with C/C++. In this paper, we describe the state of adoption of Julia in HEP,
where momentum has been gathering over a number of years.

HEP-oriented Julia packages can already read HEP's
major file formats, including TTree and RNTuple. Interfaces to some of
HEP's major software packages, such as Geant4, are available
too. Jet reconstruction algorithms in Julia show excellent performance. A number
of full HEP analyses have been performed in Julia.

We show how, as the support for HEP has matured, developments have benefited
from Julia's core design choices, which makes reuse from and integration with
other packages easy. In particular, libraries developed outside HEP for
plotting, statistics, fitting, and scientific machine learning are extremely
useful.

We believe that the powerful combination of flexibility and speed, the wide
selection of scientific programming tools, and support for all modern
programming paradigms and tools, make Julia the ideal choice for a future
language in HEP.}

\begin{document}

\maketitle

\section{Programming Languages in High-Energy Physics}
\label{sec:introduction}

\subsection{HEP Needs}

High-energy physics (HEP) is a large field, consisting of tens of thousands of
researchers, almost all of whom will need to interact with software and
contribute to software projects during their careers~\cite{2024EPJWC.29505023M}.
It is also one of the biggest, if not the biggest, generators of scientific
datasets today, with exabytes of storage used by the LHC
experiments~\cite{Collaboration:2904204}. This data is processed by a huge
corpus of software, estimated to be many tens of millions of lines in
C++ and Python, as well as Fortran, CUDA and others~\cite{hsfcwp}.

This brings a challenge for HEP software. From the point of view of \emph{code
efficiency} we require fast execution, high throughput, and
scalability at large heterogeneous computer centres and across distributed infrastructures.
Considering \emph{human efficiency} we would like a low barrier to entry for
newcomers, the ability to prototype code rapidly, a broad ecosystem of well
maintained packages, and excellent tooling for developers. These features are
needed to make software able to deal efficiently with huge datasets, as well as
accessible to a large group of developers. Stability and code preservation are also needed.

\subsection{From Fortran to the C++/Python Era}

In response to changing technology and needs the programming
languages that are dominant in HEP have evolved over time.
From~\cite{pivarski2022} we can identify three major shifts. 


\begin{description}
    \item[From assembler to Fortran] c.~1960 As computers developed from early
    specialised behemoths improved programming
    languages became available. For technical computing Fortran was the most
    effective language and HEP quickly adopted it as it brought a much improved
    syntax, as well as hardware portability.
    \item[From Fortan to C++] c.~2000 Although Fortran had many advantages, HEP had to
    develop language extensions to introduce missing concepts, such as more advanced data
    structures~\cite{Zoll:2296399}. A language which offered native object
    orientation was an attractive choice. In addition, the gap afforded by the
    end of the LEP accelerator and the construction of the LHC gave the field
    the time for a major language shift.
    \item[The rise of Python] c.~2010 Python earned a well deserved reputation as an
    excellent language for programming efficiency, and also gained ground
    through being the de facto interface to many machine learning libraries. It
    has become widely used in HEP as a complement to C++.
\end{description}

The current situation for HEP is that C++ and Python are now both widely used,
with each bringing specific advantages, as well as drawbacks. Good C++ excels at
runtime efficiency, but is a difficult language to learn, far less to master, as well as
suffering from memory safety issues and being difficult to compose. Python is
expressive, much easier to work with, is safer with memory and composes better
(via duck typing). However, it is very slow compared to C++, so not suitable for
high throughput computing.

As discussed at length in~\cite{eschle2023potential}, using two languages is not ideal: it
requires additional expertise, necessitates reimplementation of code for
performance, and reduces code reusability.

\section{Julia}

\subsection{Julia's Motivations}

The Julia programming language was
\href{https://julialang.org/blog/2012/02/why-we-created-julia/}{announced} in 2012
listing a series of ambitious goals for the language, and representing
a view into the core developers' mindset, formalised in a later
papers~\cite{bib:julia_freshapproach,10.1145/3276490}.

Julia provides a syntax as productive as Python, especially for numerical work,
whilst leveraging JAOT\footnote{Just-Ahead-Of-Time: meaning compilation to machine code happens, like C++, but just before a code block is executed, unlike C++.Julia can also cache these results between runs.} compilation to provide
\href{https://julialang.org/benchmarks/}{speed} similar to compiled, statically
typed languages like C/C++ and Rust. It utilises type inference to allow coding in a
``gradually typed'' (\S\ref{sec:juliatypes}), generic programming style, although the language will
always track types for performance behind the scenes, and types can be specified
explicitly if required. Like MatLab and Fortran, Julia's native operations and
type system support arrays as first-class entities, of any dimension, allowing
array-oriented code to be productively generated and efficiently executed, with
operations naturally ``broadcast'' element- or dimension-wise. This allows Julia
to also be effective for writing, e.g., linear algebra heavy code. In addition,
influence from the R community provides a wealth of statistical packages, and an
R-flavoured approach to plots and visualisation. As with most languages of the
21st century, Julia is a fully open-source language, with its entire codebase
freely available (and almost all of it written in Julia itself).

\subsection{Julia in Practice}
\label{sec:juliainpractice}

Julia's syntax is familiar to programmers conversant with programming performant
code in Python with NumPy, except that whitespace is not relevant
(code blocks end with \texttt{end}).

For example, the listing in Figure~\ref{code:madelbrot} shows some toy code to
generate a grayscale image of the famous Mandelbröt set.

\begin{figure}[!ht]
\centering
\begin{minted}[linenos]{julia}
using Images

function mandel(z; maxiter=80)
    c = z
    for n = 1:maxiter
        if abs2(z) > 4
            return n-1
        end
        z = z^2 + c
    end
    maxiter
end

set = [ mandel(complex(r,i)) for i=-1.:.01:1., r=-2.0:.01:0.5 ]
img = Gray.(set ./ 80)
\end{minted}
\caption{Julia implementation of the Mandelbr\"{o}t set}
\label{code:madelbrot}
\end{figure}

Here we demonstrate importing of packages (line 1); function declaration with
implicit types, positional and named parameters (line 3); an explicit loop (line
5); branch constructs (line 6); the implicit Julia use of return values (line
11); implicit loops via ``comprehensions'' (line 14) to generate a  value for
every point in an implicitly defined array; and, finally, transparent
broadcasting of operations over that array (line 15, where both the function
call to \texttt{Gray}, and the division operation are distributed over the whole
array). Unlike Python, explicit loops are optimised efficiently, and are no
slower than comprehensions or ``functional-style'' iterators (which are also
supported by Julia).

\texttt{juliaup} allows seamless management of multiple Julia releases on the same
machine, including tracking particular patch releases, and choosing the system
default.

The integrated package management in Julia, via the \texttt{Pkg} library, tracks and
maintains the dependency graph of a Julia project. State is entirely stored
within two human-readable files -- \texttt{Project.toml} (the direct
dependencies of the project) and \texttt{Manifest.toml} (the exact
resulting environment, including secondary dependencies and exact versions).
One can easily reproduce the exact environment used by
a codebase as long as these files are provided.

The Julia Package ``General Registry'' indexes packages and their releases, and
(like Rust and Javascript) relies on a public index hosted on GitHub. A help
environment allows interactive help on any keyword or symbol known a the
Read-Evaluate-Print-Loop (REPL) environment (which would be very familiar to
Python users); it is trivial to add documentation to a function or type by
simply prepending its definition with a triple-quoted docstring.

Leveraging the fact that Julia is JAOT-compiled, and that this allows its entire
standard library to be written in idiomatic Julia, the REPL also provides a
series of powerful macro utilities for inspecting the byte- and machine-code
generated for a given expression (\verb$@code_lowered$, \verb$@code_native$) and
for locating and displaying the source code for any function in the current
namespace, including from the standard library (\verb$@less$). Profiling of code
in the REPL is similarly directly supported via macros
(\href{https://github.com/JuliaCI/BenchmarkTools.jl}{\texttt{@benchmark}}),
which provide detailed performance sampling. Extended introspection and
profiling tools are also available in optional packages, such as
\href{https://github.com/tecosaur/About.jl}{\texttt{About.jl}}, providing
information on memory layout of datatypes and thread safety of functions.

A well-supported VSCode extension is available for the language, which also
supports the standard Language Server Protocol allowing it to support
development in other editors. This includes the usual benefits of code
completion, including using Copilot and other AI extensions, Unicode completion,
linting and highlighting. There is a built-in interactive debugger, as well as
innovative tools to break into a REPL in the middle of code execution
(\href{https://github.com/JuliaDebug/Infiltrator.jl}{\texttt{Infiltrator.jl}}).

Julia is now popular enough to have many online training materials, an active
support community on Slack and Discourse, and many postings on Stackoverflow.

Finally, Julia is one of the founding languages supported by Jupyter - being the
``Ju'' in the portmanteau -- and also provides other native notebook
implementations such as \texttt{Pluto.jl}. 

\subsection{Key Design Features for Performance}

Whilst there are many features of the Julia language design which contribute to its 
performance, and productivity, we highlight two particular ones here.

\subsubsection{Type System}
\label{sec:juliatypes}

Julia's type system is an expressive, but simple, tree of sets, where only leaves
of the tree can be instantiated as concrete types. The higher levels of abstract
types provide a means for expressing categories of types which are all supported
by an operation, all the way up to the \verb$Any$ type, which contains all other
types, and is the default type for function parameters if none are specified.
For example, using Julia's notation for ``is a subtype of'', \texttt{<:}, the concrete type
\verb$Float64$ is in the following hierarchy:

\begin{minted}{julia}
Float64 <: AbstractFloat <: Real <: Number <: Any
\end{minted}

All higher levels are abstract, and thus cannot be used as the type for a
binding. Other sibling subtypes of \verb$Real$ include \verb$Float32$,
\verb$Int64$ and \verb$BigInt$.

Abstract types \textit{can} be used as the type of an element of a container
type, resulting in a container with boxed types, where each element can be any
concrete subtype of the element; and similarly for function parameters, where
they constrain that parameter to taking the relevant concrete subtypes as
values. 

Together, this allows complex type expressions such as

\begin{minted}{julia}
AbstractArray{T,2}
\end{minted}

This represents a function argument that takes any Array-like type, of 2
dimensions, with an arbitrary element type \texttt{T} (which could be further constrained
by annotation). As the subtypes of \verb$AbstractArray$ include performance
specialised cases like diagonal and sparse arrays, as well as accelerated cases, like
GPU arrays (computed on an accelerator, not the CPU), then all of these cases are
naturally handled by the same function that supports this one type. 

\subsubsection{Multiple dispatch}

The type system above allows for efficient specialisation of functions by the
JAOT backend, based on the types they are called with. Unusually, Julia exposes
this to the user, allowing specialisations of functions to be dispatched on the
type of any (or all) parameters it is called with, not merely the first. This
\textit{multiple dispatch} allows effective provision of efficient special cases
for functions to be provided, powered entirely by the type system.

Extending the nomenclature for the \textit{single dispatch} used by
class-based object orientation, Julia considers the variants of a function
distinguished by their type signatures to be separate ``methods''.

The runtime resolution of function dispatch also allows seamless composition
of functionality between packages, without the need for special case
code in any of the involved.

For example, in the code listing from Figure~\ref{code:plotmeasure} below, the
\texttt{Plots} and \texttt{DifferentialEquations} packages do not know about
\texttt{Measurements} -- and yet the types provided by \texttt{Measurements} can
seamlessly extend their functionality (as well as providing their own extensions
of mathematical operations), resulting in the plot combining all the features of
the packages seamlessly.

\begin{figure}[!ht]
\begin{subfigure}[t!]{0.40\textwidth}
    
    \begin{minted}[fontsize=\footnotesize]{julia}
    using Measurements, Plots
    using DifferentialEquations

    g = 9.79 ± 0.02; 
    L = 1.00 ± 0.01;
    u0 = [0.0 ± 0.0, π \ 60 ± 0.01]
    tspan = (0.0 ± 0, 1.0 ± 0) 
    function p(du,u,p,t)
        du[1]=u[2]
        du[2]=-(g/L)*u[1]
    end
    prob = ODEProblem(p, u0, tspan) 
    sol = solve(prob, Tsit5());
    plot(sol.t, getindex.(sol.u,2)) 
    \end{minted}
\end{subfigure}
\hfill
\begin{subfigure}[t!]{0.55\textwidth}
    \includegraphics[width=0.95\textwidth]{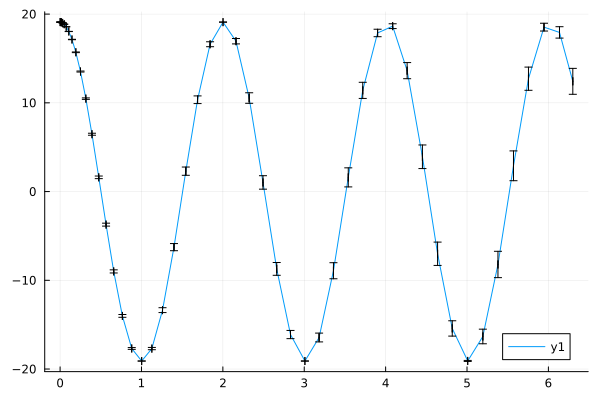}
\end{subfigure}
    \caption{Plot with \texttt{Measurements} of numerical solution using \texttt{DifferentialEquations}, 
    composed via multiple dispatch and the Julia type system.}
    \label{code:plotmeasure}
    \end{figure}

\section{Julia for Scientific Computing}
\label{sec:juliascicomp}
    
\subsection{GPU Programming}
\label{sec:gpuprog}

Julia's JAOT compilation model makes it ideal for running on GPUs. Julia
supports GPU programming for specific backends such as
\texttt{CUDA.jl}~\cite{besard2018juliagpu} for NVIDIA,
\texttt{AMDGPU.jl}~\cite{julian_samaroo_2025_14826765} for AMD,
\href{https://github.com/JuliaGPU/Metal.jl}{\texttt{Metal.jl}} for M-series Mac
devices, and \href{https://github.com/JuliaGPU/oneAPI.jl}{\texttt{oneAPI.jl}}
for Intel devices. They require a minimal amount of development and support
writing kernels with granular control.

Array based calculations are trivial to execute on the GPU. In the listing in
Figure~\ref{code:arraysgpu} an array is copied to the GPU and then executes a
trivial operation, but mainly shows that there is basically no boilerplate required to access GPU computations. 

\begin{figure}[!ht]
\begin{center}
\begin{minted}{julia}
using CUDA

a = CuArray([1,2,3,4])
a * 2
\end{minted}
\caption{Array based calculations for the \texttt{CUDA.jl} backend - the code for other backends is essentially the same.}
\label{code:arraysgpu}
\end{center}
\end{figure}

It is not always possible to write an algorithm and run it using high-level
array abstractions. Instead one has to write a GPU kernel -- Kernel
programming is close to the native toolkits.  Writing and maintaining specific
backend code is undesirable, from a portability point of view. Therefore Julia
also supports a generic backend,
\texttt{KernelAbstractions.jl}~\cite{BESARD201929,8471188}. This allows
developers to separate the mathematical logic of their codes from the specific
GPU backend on which it will run, resulting in high code portability. This
feature is used heavily in Julia codes from \texttt{Flux.jl}~\cite{innes:2018},
for machine learning; to
\texttt{Oceananigans.jl}~\cite{Ramadhan_Oceananigans_jl_Fast_and_2020}, a fluid
dynamics code.

Memory management on the GPU is
\href{https://cuda.juliagpu.org/stable/usage/memory/}{mostly automated},
including garbage collection of unreachable objects, but more fine grained
steering is available.

\subsection{Julia HPC codes}
\label{sec:julia-hpc}
Beyond the single-node practices exemplified above, large-scale applications
often require high-performance computing (HPC) technologies. Here Julia has
established itself as a strong contender alongside well-known HPC-capable
programming languages such as C, C++, and Fortran. Benchmarks evaluating intra-
and inter-node communication on CPUs have shown that Julia introduces negligible
overhead compared to native C implementations~\cite{hunold2020benchmarking}. 

A particularly noteworthy achievement is the near-zero-loss integration of MPI
through \texttt{MPI.jl} \cite{byrne2021mpi}, enabling efficient parallel
communication. Furthermore, when it comes to computational performance, Julia
has demonstrated its ability to keep up with traditional HPC technologies. For
instance, in single-CPU-node scenarios, Julia's performance matches that of
vendor-optimised libraries \cite{giordano2022productivity}. Similarly,
performance portability studies in multi-node CPU/GPU benchmarks confirm Julia's
competitiveness across architectures
\cite{teichgraber2022julia,godoy2023evaluating}.

Beyond controlled benchmarking environments, Julia has also proven its
capabilities in real-world HPC applications. One prominent example is the
\texttt{Celeste.jl} project \cite{RiegerIEEE:2018,RiegerAAS:2019}, where a peak
performance of \(1.54\,\mathrm{PFLOPs}\) was shown, demonstrating Julia's
scalability, efficiency in large-scale computations and, consequently,
membership of the petaflop club.

More recently, Julia has been advancing multi-GPU and multi-CPU applications in
large-scale simulations. Notable examples include
\href{https://github.com/PTsolvers/FastIce.jl}{\texttt{FastIce.jl}} for
high-resolution flow simulations and various geocomputing applications
leveraging HPC-related packages like
\href{https://github.com/omlins/ParallelStencil.jl}{\texttt{ParallelStencils.jl}}
and
\href{https://github.com/eth-cscs/ImplicitGlobalGrid.jl}{\texttt{ImplicitGlobalGrid.jl}}.
These packages enable high-performance stencil computations and efficient
distributed memory parallelism, reinforcing Julia’s role as a modern HPC
language.

\section{Julia in HEP}

\subsection{Challenges}

High-energy physics faces significant challenges in software efficiency and
scaling for the years to come, particularly driven by the physics programme of
the high-luminosity LHC~\cite{hsfcwp}. Data volumes will rise sharply and data
complexity will also
increase~\cite{CERN-LHCC-2022-005,Software:2815292,Valassi2021}. Therefore
software efficiency and scaling are paramount. As we have seen in
\S~\ref{sec:julia-hpc}, the Julia language can achieve very efficient
execution on HPC clusters of CPUs and GPU accelerated nodes. It has also
\href{https://indico.cern.ch/event/1410341/contributions/6135602/}{been shown}
that Julia artefacts can be effectively distributed on CVMFS, so that efficient
running on, e.g., WLCG sites, would be possible.

HEP has millions of lines of legacy code, with which any new language must
interoperate. In this respect Julia is in a very strong position. For calling
into C or Fortran, Julia offers a
\href{https://docs.julialang.org/en/v1/manual/calling-c-and-fortran-code/}{simple
\texttt{@ccall}} macro, which is a direct, no overhead, no boilerplate interface
to libraries compiled from these languages. There are many existing examples of
calls to these foreign libraries being used in Julia, e.g., at a low level Julia
itself uses the OpenBLAS library~\cite{6877458}, written in C and modern
Fortran. 


C++ code is more difficult to interface to (a `feature' of C++ itself), and the
smoothest way to achieve it is to write a small wrapper in C++ to interface to
Julia, making use of the Julia interface package
\href{https://github.com/JuliaInterop/CxxWrap.jl}{\texttt{CxxWrap.jl}}. This
process can be automated with the helper program
\href{https://github.com/grasph/wrapit}{WrapIt}, which can generate wrapper code
automatically from C++ headers. With this, wrapping very large libraries becomes
much easier, as shown below in \S~\ref{sec:simulation}.

The other major desiderata for a programming language is that it is efficient
for programmers -- and for HEP this must cover the spectrum from novice coders
to experienced software engineers. As illustrated in \S\ref{sec:juliainpractice} Julia's development ecosystem is optimised for
efficient code development, leading to widespread adoption in
science~\cite{perkel-julia-science}.

Of course, a language having suitable \emph{general} features, does not
automatically mean that everything needed by researchers in a particular field
have the tools and packages that they need. In the next sections we outline the
growing ecosystem of packages specifically developed for HEP that make it
possible to be productive with Julia from day 1 of coding. 

\subsection{HEP Data Formats}
\label{sec:hepdataformats}

Most HEP experimental data is stored in ROOT~\cite{ROOT:2011zz} format, making
it crucial for Julia to have the ability to read and write data in this format.
Several options are available for achieving this functionality. One possibility
is to create Julia bindings to the ROOT C++ package. However, this approach
presents technical challenges, as it requires cross-compilation to generate
binary artifacts, a process that is not yet fully supported by ROOT. Despite
this, progress is being made with the \texttt{ROOT.jl}~\cite{ROOT_jl} package,
which aims to address these challenges providing a direct interface to ROOT's C++ code.

An alternative is the \texttt{UnROOT.jl}~\cite{UnROOT:2022vnn} package, which
offers a pure Julia implementation of the ROOT I/O format, independent of ROOT
or Python, thus implying a lightweight solution, with far fewer dependencies and
a light. This package supports transparent reading of both TTree and the newer
RNTuple storage formats. While writing functionality is still under development,
\texttt{UnROOT.jl} provides fast, memory-efficient, and lazy data access,
reading only the necessary parts of a file on demand.

The \texttt{UnROOT.jl} package can also be used to read EDM4hep
\cite{EDM4hep:2022leb} event data, in combination with the
\href{https://github.com/JuliaHEP/EDM4hep.jl}{\texttt{EDM4hep.jl}} package. The
EDM4hep event data model aims to establish a standard for storing and exchanging
event data in HEP experiments. The \texttt{EDM4hep.jl} package provides a pure
Julia interface to this model, generating user-friendly Julia types that map to
the EDM4hep data structure. These types are compatible with both arrays of
structures and structures of arrays when reading data files typically produced
by C++ programs. This structured approach makes data access more intuitive
compared to reading flat n-tuples, facilitating the development of analysis code
in a more natural and efficient way.

\subsection{Event Generators}

The numerical calculation of amplitudes for hard scattering processes and Monte
Carlo event generation plays a fundamental role in high-energy physics analysis
workflows, with a long history (\cite{campbell2024event}). Consequently,
integrating various aspects of physics models with specialised numerical
algorithms in a high-performance computing environment is recognised as one
major challenge in HEP software development for future HEP experiments
(\cite{HEPSoftwareFoundation:2017ggl, HSFPhysicsEventGeneratorWG:2020gxw,
HSFPhysicsEventGeneratorWG:2021xti}) and beyond.

With the open-source framework \texttt{QuantumElectrodynamics.jl}
\cite{qedjl-github}, the first steps in exploring how Julia
can address these challenges are taken. Specifically, the framework
facilitates the numerical calculation of scattering amplitudes and the
implementation of Monte Carlo event generation within the domain of perturbative
and strong-field quantum electrodynamics (cf.~\cite{Fedotov:2022ely}). The
overall structure of the framework is unified through interfaces defined in
\texttt{QEDbase.jl}, which provides standardised representations for fundamental
mathematical objects: four-momenta, bi-spinors, and phase space points.
Additionally, these interfaces extend to the configuration entities: 
scattering processes, computational models, and phase space layouts. Further
interfaces are provided for computable quantities, such as differential
cross-sections, and various samplers for Monte Carlo event generation. Here,
Julia’s multiple dispatch mechanism proves particularly powerful, allowing
different implementations to be seamlessly integrated without explicitly
specifying types within the interface definitions. Moreover, the domain-specific
language facilitated by these interfaces -- covering elements such as processes,
models, and phase space layouts -- combined with multiple dispatch enables the
straightforward incorporation of analytical formulas whenever they are
available. This typically reduces to implementing a single method for a specific
function signature.

Beyond its interface definitions, \texttt{QuantumElectrodynamics.jl} provides
concrete implementations for all major components: \texttt{QEDcore.jl} handles
the fundamental mathematical structures, \texttt{QEDprocesses.jl} computes
differential cross sections, and \texttt{QEDevents.jl} offers different samplers
for event generation. The \texttt{QEDFeynmanDiagrams.jl} package, as part of
\texttt{QEDprocesses.jl}, facilitates the calculation of scattering amplitudes
for arbitrary QED processes by leveraging Julia's metaprogramming and code
generation capabilities. Built on top of \texttt{ComputableDAGs.jl}, the
generated code can be directly analysed and manipulated within Julia itself,
even without leaving the session. This enables meta-optimizations based on
domain-specific knowledge, such as recognising patterns in Feynman diagrams and
exploiting mathematical properties like distributivity.

Finally, by leveraging multiple dispatch and various array abstractions from the
Julia ecosystem (\S\ref{sec:gpuprog}) the generated
code can be seamlessly compiled and executed on GPUs as well as CPUs,
without requiring any modifications to its underlying structure. While some
framework components are still under development and will be further extended to
address broader challenges in high-energy physics, the initial structures
demonstrate great potential. The modular design, combined with Julia's
capabilities, has the potential to contribute to future developments in
numerical calculations and event generation within HEP.

\subsection{Simulation}
\label{sec:simulation}

Detector simulation is a crucial component of every HEP
experiment, playing a key role both during the design and conception of the
detector and later in data analysis. The most widely used toolkit for this
purpose is Geant4~\cite{GEANT4:2002zbu}, a C++-based framework with over 2
million lines of code.

Given its complexity and extensive adoption, a complete rewrite of Geant4 in a
new language is highly impractical. Instead, this presents an opportunity to
explore Julia's interoperability with other languages. One particular challenge
arises from Geant4's callback-based user interface, which relies on C++ virtual
methods invoked at specific points during particle transport. Application
developers must implement these callbacks to configure and control the
simulation and extract relevant data. However, integrating this
mechanism in Julia is more complex than in other languages, as Julia does not
natively support virtual methods.

The \href{https://github.com/JuliaHEP/Geant4.jl}{\texttt{Geant4.jl}} package has
been developed to provide a Julia interface to Geant4~\cite{ALLISON2016186}. It
leverages
\href{https://github.com/JuliaInterop/CxxWrap.jl}{\texttt{CxxWrap.jl}}, a
package that enables calling C++ functions and types from Julia. Similar to
Python's static bindings, invoking C++ code from Julia requires explicit wrapper
definitions for each method exposed to Julia. However, given Geant4's large and
complex codebase, manually writing and maintaining these wrappers is not a
viable approach, especially for making it more sustainable with future toolkit
updates. To address this, we use
\href{https://github.com/grasph/wrapit}{WrapIt}, which automates wrapper
generation by utilising the Clang library to parse C++ header files and extract
class declarations. This automation significantly reduces development effort and
ensures long-term maintainability of the interface.

Integrating Geant4 with Julia allows researchers to take advantage of Julia's
high-level programming capabilities and performance (e.g., to plot results or code user actions), while retaining the
full functionality and efficiency of the underlying Geant4 toolkit. This integration also
provided an opportunity to rethink and streamline the interface, making it more
intuitive and user-friendly. The package interface ensures that
application developers can concentrate on the essential aspects of their
simulations while minimising configuration overhead. Boilerplate code and C++
idiosyncrasies are hidden, allowing for a cleaner, more concise approach to
defining simulations. The performance of the \texttt{Geant4.jl} package is
comparable to that of the native C++ Geant4 toolkit, demonstrating the
feasibility of using Julia for HEP detector simulation.

\subsection{Reconstruction}

Reconstruction of HEP data is a core task for the field, and a huge amount of
software, often rather detector specific, is dedicated to that task. Even for
very detector specific code there are some early demonstrations that Julia would
be suitable, and can be competitive with modern optimised C++ for a large
experiment on multiple different GPU backends, e.g., for the CMS pixel
detector~\cite{patatrack_julia}. For smaller experiments, there are already
end-to-end solutions based on Julia, as shown below in
\S\ref{sec:e2dlegend}.

For generic reconstruction tasks there is less code in general, however, for the
task of sequential jet reconstruction, the Fastjet
package~\cite{Cacciari:2011ma} has seen very wide adoption across many
experiments. There is now a native Julia reimplementation of many of the same
algorithms as are found in Fastjet, in the
\href{https://doi.org/10.5281/zenodo.12671414}{\texttt{JetReconstruction.jl}}
package~\cite{polyglot-jets-chep23}. Code ergonomics have been found to be
superior, taking advantage of many of Julia's features such as array
broadcasting and generic programming. Performance is, on average 16\% better
than Fastjet for popular $pp$ reconstruction algorithms, and 33\% better for the
Durham algorithm for $e^+e^-$ events in FCCee
studies~\cite{fast-jet-reco-julia}. Using Julia's package extension mechanisms,
the ubiquitous EDM4hep~\cite{EDM4hep:2022leb} event data model is
supported directly. This adds to the user experience, where not only can the
package offer simpler, more generic interfaces than Fastjet, but there is the
ability to seamlessly hook into the rest of the Julia ecosystem, e.g., for jet
visualisation with the native Julia \texttt{Makie.jl}
package~\cite{Danisch2021}.

\subsection{Analysis}

In most HEP experiments, after the main production of reconstructed objects at a
level suitable for analysis, the path to publication diverges into many
individual and group analyses. Here, as there is a relative independence from
code in other languages, Julia can be a highly efficient alternative to other
analysis pipelines. An overview of the suitability of Julia for analysis is
found in Stanitzki and Strube~\cite{Stanitzki:2020bnx}. Statistical tools are
well supported in Julia, e.g., the \href{https://github.com/JuliaStats/}{Julia Statistics
community} is a good entry point. HEP
specific histograming needs are supported by
\href{https://github.com/Moelf/FHist.jl/}{\texttt{FHist.jl}}, which provides error
propagation and high performance.

At this phase of the analysis, data, usually in ROOT format, is read
(\S\ref{sec:hepdataformats}). Then there is a growing ecosystem of packages
dedicated to HEP analysis. For example, for hadron physics one can use
\href{https://github.com/mmikhasenko/ThreeBodyDecays.jl}{\texttt{ThreeBodyDecays.jl}}
to build hadronic decays with cascade reactions. Partial-wave analysis is then a
common technique, which is supported by
\href{https://github.com/mmikhasenko/PartialWaveFunctions.jl}{\texttt{PartialWaveFunctions.jl}},
as well as fitting hardonic line shapes, via
\href{https://github.com/mmikhasenko/HadronicLineshapes.jl}{\texttt{HadronicLineshapes.jl}}.

Such a corpus of packages have been used to support many final physics
analyses, e.g.,~\cite{Aaij2022,PhysRevD.104.L091102,Bibrzycki2021}.

\subsection{End-to-end Computing}
\label{sec:e2dlegend}

The LEGEND~\cite{LEGEND:2017AIPC} experiment demonstrates how Julia can be used
as a basis for end-to-end analysis in a larger physics experiment with multiple
subsystems. LEGEND uses Julia as its official secondary software stack, both to
verify the results of the primary software stack (written in Python) and as a
test-bed for future software technologies. The whole data analysis chain,
encompassing raw waveform data signal processing, ML-based data quality cuts,
data calibration, event building and high-level statistical analysis is
implemented completely in Julia. LEGEND uses the Bayesian Analysis
Toolkit in Julia (\texttt{BAT.jl})~\cite{Schulz:2021BAT} as its primary Bayesian
framework for both background decomposition and final physics analysis, and the
Julia package \texttt{SolidStateDetectors.jl}~\cite{Abt:2021SSD} for detector simulation
and detector design. With the exception of a custom Geant4-based software, the
LEGEND collaboration is now able to perform any simulation and analysis task in
Julia.

\section{Conclusions}

As we have shown Julia is an ideal language for scientific computing. It
combines runtime efficiency and scalability, which is as good as C++ and
Fortran, with a modern high-productivity setup, suitable for rapid prototyping.
AI and ML packages are supported, either as wrapped versions of Python/C++, or as native implementations.
The higher level features of the language lend themselves to compact code that
expresses mathematical operations in a natural way. Multiple dispatch, supported
by Julia's type system, allows excellent code composition, with a very high
degree of code reuse. Julia's package system allows for easy setup of
environments and excellent reproducibility. The language also interfaces with
existing HEP code bases, allowing reuse of large codes, with improved ergonomics
and integration with other Julia packages.

The Julia community in HEP is young, but there is a growing HEP specific set of
packages that are making Julia productive and useful specifically for
high-energy physics. The advantages of the language we find to be compelling and
a path of gradual adoption, helped by
\href{https://hepsoftwarefoundation.org/activities/juliahep.html}{groups like
JuliaHEP} is both possible and desirable.

\sloppy
\raggedright
\bibliography{julia-in-hep}

\end{document}